\documentclass[aip,
 amsmath,amssymb,
 preprint,%
floatfix]{revtex4-1}

\usepackage{graphicx}
\usepackage{dcolumn}
\usepackage{bm}
\usepackage{enumitem}
\newlist{steps}{enumerate}{1}
\setlist[steps, 1]{label = Step \arabic*:}
\usepackage{listings}
\usepackage[utf8]{inputenc}
\usepackage[T1]{fontenc}
\usepackage{mathptmx}
\newcommand{\bra}[1]{\ensuremath{\left\langle#1\right|}}
\newcommand{\ket}[1]{\ensuremath{\left|#1\right\rangle}}

\newcommand{\bracket}[2]{\ensuremath{\left\langle #1 \middle| #2 \right\rangle}}
\begin{document}

\preprint{}

\title[]{Non-adiabatic Dynamics in a Laser Field with Floquet Fewest Switches Surface Hopping: The Need for An Accurate Treatment of Coherence and Decoherence Remains}

\author{Zeyu Zhou}
 
\author{Hsing-Ta Chen}

\author{Abraham Nitzan}%

\author{Joseph Eli Subotnik}
 \email{subotnik@sas.upenn.edu}
 \affiliation{Department of Chemistry, University of Pennsylvania, Philadelphia, Pennsylvania 19104, United States}

\date{\today}

\begin{abstract}
We investigate two well-known approaches for extending the fewest switches surface hopping (FSSH) algorithm to periodic time-dependent couplings. The first formalism acts as if the instantaneous adiabatic electronic states were standard adiabatic states, which just happen to evolve in time. The second formalism replaces the role of the usual adiabatic states by the time-independent adiabatic Floquet states. For a set of modified Tully model problems, the Floquet FSSH (F-FSSH) formalism gives a better estimate for both transmission and reflection probabilities than the instantaneous adiabatic FSSH (IA-FSSH) formalism. More importantly, only F-FSSH predicts the correct final scattering momentum. Finally, in order to use Floquet theory accurately, we find that it is crucial to account for the interference between Floquet states. Our results should be of interest to all those interested in laser induced molecular dynamics.
\end{abstract}

\maketitle

\section{\label{sec:one}Introduction}
Electronic non-adiabatic transitions and photon excitations are among the most important dynamical processes in the field of spectroscopy and photochemistry, and while usually considered separate processes, these phenomena can both enhance or compete with each other. 
Specifically, electrons in a molecular system interacting with an incident laser field can transition between adiabatic states (i) through the non-adiabatic coupling beyond the Born--Oppenheimer approximation or (ii) through the radiative coupling in conjunction with absorption or emission of photons.
Over the past decades, many exciting phenomena, ranging from molecular photodissociation\cite{regan1999ultraviolet,franks1999orientation, corrales2014control, hilsabeck2018nonresonant, hilsabeck2019photon} to coherent X-ray diffraction,\cite{glownia2016self,fuller2017drop} have highlighted the importance of the dynamical interplay between non-adiabatic transitions and photoexcitation processes.
These recent observations demonstrate a need for accurate theoretical treatments of non-adiabatic molecular dynamics as driven by an external time-dependent field.
However, while exact dynamics may be obtained for specific idealized models, most practical simulation schemes for large, realistic systems rely on a mixed quantum--classical framework---treating the electronic degrees of freedom with quantum mechanics and describing the nuclear motions and the laser field using classical mechanics/electrodynamics. 

Given the success of semiclassical methodologies for studying non-adiabatic molecular dynamics over the past 30 years, the most natural strategy to capture laser-driven non-adiabatic dynamics is to generalize these semiclassical methods to treat radiative coupling in a similar fashion.\cite{marquetand2016challenges, bajo2012mixed} 
Today, several different approaches based on Tully's fewest switches surface hopping (FSSH) have been proposed,\cite{tully1990molecular} including the surface hopping including arbitrary couplings (SHARC)\cite{richter2011sharc,mai2018nonadiabatic} and field-induced surface hopping (FISH)\cite{mitric2011field,lisinetskaya2011simulation} approaches. Each of these approaches applies a modified hopping mechanism to account for the interaction with the laser field. 
Within these methods, the electronic Hamiltonian is expressed in terms of the classical parameters of the nuclear coordinates and the incident laser field that explicitly depends on time; 
diagonalizing the electronic Hamiltonian yields the instantaneous adiabatic states and potential energy surfaces (PESs).\cite{thachuk1996semiclassical, thachuk1998semiclassical}
The basic premise of SHARC is to modify Tully's FSSH algorithm with minimal changes to FSSH such that effectively one runs Tully's algorithm on top of an instantaneous adiabatic representation: the electronic wavefunction evolves according to the Schr\"odinger equation with the time-dependent Hamiltonian and a swarm of classical trajectories is propagated along the instantaneous adiabatic PESs. 
The active PES of each classical trajectory can change from one to another through a hopping event associated with (i) the nuclear-derivative (non-radiative) coupling (which should in theory conserve the total energy of the system) or (ii) the time-derivative (radiative) coupling which allows the electronic system to absorb/emit radiation energy and which does not conserve energy.
These so-called instantaneous adiabatic FSSH methods (IA-FSSH) have been used to study gas phase photodissociation\cite{richter2011sharc} and optimal control of the trans-cis isomerization.\cite{mitric2009laser}

Now, despite its growing popularity, IA-FSSH does suffer several drawbacks. 
First, the FSSH problem of decoherence/recoherence is expected to be exacerbated for driven non-adiabatic dynamics. 
Since the instantaneous PESs oscillate rapidly at the frequency of the laser field, the electronic coherence becomes more complicated and difficult to recover with FSSH (which does not include any scheme to deal with the repeated separation and recombination of wave packets moving on different PESs).\cite{horenko2001theoretical}
While several decoherence strategies exist,\cite{schwartz1994aqueous, bittner1995quantum, schwartz1996quantum, prezhdo1997evaluation, prezhdo1997mean, fang1999improvement, fang1999comparison, volobuev2000continuous, hack2001electronically, wong2002dissipative, wong2002solvent, horenko2002quantum, jasper2005electronic, bedard2005mean, subotnik2011decoherence, subotnik2011new, landry2012recover, mai2018nonadiabatic, mignolet2019excited} one might expect fatal problems for driven IA-FSSH non-adiabatic dynamics because recoherence effects cannot be captured by non-interacting FSSH trajectories.\cite{petit2014howto,petit2014calculating, miao2019revisiting, martens2016surface}
Second, IA-FSSH treats the absorption/emission of radiation energy as a continuous phenomenon and ignores all quantum features of light. Thus, radiation-induced hopping can occur at any energy difference between two PESs, usually without regard to energy conservation, such that one might expect to find not the most accurate description of energy absorption/emission. One phenomenological means to solve this problem is to allow the radiative hopping only within a restricted bandwidth of the photon energy, but it is unclear how effective this approach will be.

In very recent years, Floquet FSSH (F-FSSH) method has emerged to be an appealing alternative for simulating laser-driven non-adiabatic dynamics, especially for a time-periodic field.\cite{bajo2012mixed, fiedlschuster2016floquet}
Unlike the instantaneous adiabatic representation for a periodic couplings, F-FSSH eliminates all explicit dependence on time by expanding the electronic wavefunction in a Floquet state basis (the diabatic states dressed by $e^{im\omega t}$ for an integer $m$ and the laser frequency $\omega$).
For a periodic field, the resulting Hamiltonian, the so-called Floquet Hamiltonian, is \emph{time-independent}, albeit of infinite dimension of $m$.
Thus, the Floquet quasi-energy surfaces obtained by diagonalizing the Floquet Hamiltonian are also time-independent, so that one can simply employ the standard FSSH method in the Floquet state representation with a minimal modification:
the infinite-dimensional wavefunction evolves following the Schr\"odinger equation under the Floquet Hamiltonian and classical trajectories move along Floquet quasi-energy surfaces. 
When a classical trajectory passes through a crossing of two Floquet quasi-energy surfaces, the active surface of the classical trajectory can change by Tully's stochastic hopping mechanism while conserving the Floquet quasi-energy.
Furthermore, the F-FSSH method can account for quantized photon absorption/emission because, by construction, the Floquet quasi-energy surfaces of the same diabatic state are separated by an integer multiple of photon energy ($m\hbar\omega$).

Nevertheless, despite these positive attributes, in practice, there are several difficulties when implementing the F-FSSH method for driven non-adiabatic dynamics. 
First, many crossings of Floquet quasi-energy surfaces are trivial crossings, i.e. the derivative coupling becomes infinity at the exact crossing point. 
For any FSSH simulation carried out with a finite time step, trivial crossings may be missed when propagating a classical trajectory, which leads to unphysical multi-photon excitations. 
Recently, several approaches have been developed to capture trivial crossings,\cite{hammes1994proton, fabiano2008implementation, barbatti2010non, plasser2012surface, fernandez2012identification, nelson2013artifacts,  wang2014simple, meek2014evaluation, jain2016efficient, lee2019solving} but none have been applied to the F-FSSH method; usually these methods have been applied to simulations with fewer than 20 states, and we will show that some further adjustments after are necessary. 
Second, as we will show below, reconstructing the real electronic wavefunction from the Floquet wavefunction requires an accurate treatment of the interference between wavepackets on different Floquet quasi-energy surfaces.
When summing over all the trajectories, the diabatic population includes two contributions: (i) the number of trajectories ending up on the same diabatic state and (ii) the summation over the coherences of the Floquet states associated with each diabatic state.
We are unaware of any previous discussion of this point with regard of F-FSSH in the existing literature.

With this background in mind, our goals for this papers are to rigorously investigate and compare F-FSSH with IA-FSSH and exact solutions for modified versions of two of Tully's model problems. 
This paper is arranged as follows: In section \ref{sec:two}, we review both the IA-FSSH and F-FSSH formalisms. 
In section \ref{sec:3}, we summarize the differences and modifications which we have implemented in order to construct meaningful and efficient surface hopping calculations.
In section \ref{sec:4}, two typical model problems are introduced and the results of both formalisms are compared with exact quantum calculations. 
We conclude and analyze the advantages and drawbacks of F-FSSH in section \ref{sec:5}.

\textit{Notation.} A Floquet basis requires a great number of indices; the notation can be hard to follow. In general, an arrow on the top of a letter denotes a vector. For the reader's convenience, all labels are listed in Table \ref{table:1}.

\begin{table}
\caption{\label{table:1}Notation for the Present Paper }
\begin{ruledtabular}
\begin{tabular}{c c}
variable & definition \\
\hline
$\vec{R}$, $\vec{v}$, $M$& Position, velocity and mass of nuclei\\
$\mu$, $\nu$ & Index for electronic states\\
$m$, $n$ & Index for Fourier expansion\\
$J$, $K$ & Index for Floquet states\\
$\hat{H}_{\text{tot}}$, $\hat{H}_{el}$ & Total Hamiltonian and Electronic Hamiltonian\\
$\hat{\mathbb{T}}_{\vec{R}}$ & Kinetic operator for nuclear DoF\\
$|\phi_{j}^{ad}(\vec{R}, t)\rangle$ & Instantaneous adiabatic basis of the electronic Hamiltonian \\
$c_{j}$ & Electronic amplitude on instantaneous adiabatic state i\\
$V^{el}_{j}(\vec{R}, t)$ & Eigenvalues of instantaneous electronic Hamiltonian $\hat{H}_{el}(R, t)$\\
$\vec{d}_{jk}$ & Derivative coupling \\
$T_{jk}$ & Time-derivative coupling matrix \\
$\hat{H}_{F}$ & Floquet Hamiltonian \\
$\ket{\Phi^{J}(\vec{R}, t)}$ & Electronic Floquet state $J$\\
$\hat{\cal H}^{F}$ & Floquet Hamiltonian after Fourier transform \footnote{The dimension of this transformed Floquet Hamiltonian is the dimension of the electronic Hamiltonian $\hat{H}_{el}$ times the number of Fourier modes after truncation}\\
$\tilde{c}_{J}$ & Electronic amplitude on Floquet state $(J)$\footnote{Same dimension as the Floquet Hamiltonian $\hat{\cal H}^{F}$}\\
$\varepsilon_{J}$, $\ket{G^{J}(\vec{R})}$ & Eigenvalues and eigenvectors of Floquet Hamiltonian $\hat{\cal H}^{F}$\\
$dt_{c}$, $dt_{q}$ & Time steps for nuclear and electronic propagation \\
$U_{JK}$ & Overlap matrix with elements \\
&$U_{JK}(t_{0} + dt_{c}/2) = \bracket{\Phi^{J}(t_{0})}{\Phi^{K}(t_{0}+dt_c)}$\\
\end{tabular}
\end{ruledtabular}
\end{table}

\section{\label{sec:two} Theory}
\subsection{\label{sec:2.1} Illustration of the problem}
Consider the case of a molecule illuminated with continuous wave (CW) light. The incoming light acts as a periodic coupling  $\hat{V}(t)=\hat{V}(t + T_{0})$ between electronic states with period $T_{0}$. Thus, the total Hamiltonian has the following form
\begin{eqnarray}
\hat{H}_{\text{tot}}(t)= \hat{\mathbb{T}}_{\vec{R}} + \hat{H}_{el}^{0} + \hat{V}(t) = \hat{\mathbb{T}}_{\vec{R}} + \hat{H}_{el}(t)
\label{eq:one}
\end{eqnarray}
where $\hat{\mathbb{T}}_{\vec{R}}$ is the kinetic operator and $\hat{H}_{el}^{0}$ is the time-independent electronic Hamiltonian. 
Any initial wavefunction can be propagated by the instantaneous propagator

\begin{eqnarray}
\ket{\Psi(t_{0}+dt)} = \exp\Bigl(-\frac{i\hat{H}_{\text{tot}}(t_0)dt}{\hbar}\Bigr)\ket{\Psi(t_{0})}
\label{eq:two}
\end{eqnarray}

For exact calculations in a few dimensions, the propagator can be expanded according to Tr\"{o}tter decomposition
\begin{eqnarray}
\exp\Bigl(-\frac{i\hat{H}_{\text{tot}}(t_0)dt}{\hbar}\Bigr) = \exp\Bigl(-\frac{i\hat{\mathbb{T}}_{\vec{R}}dt}{2\hbar}\Bigr)\exp\Bigl(-\frac{i\hat{H}_{el}(t_0)dt}{\hbar}\Bigr)\exp\Bigl(-\frac{i\hat{\mathbb{T}}_{\vec{R}}dt}{2\hbar}\Bigr)
\label{eq:three}
\end{eqnarray}

In practice, the wavefunction is transformed between the position and momentum representations, for which the electronic Hamiltonian and the kinetic operator are block diagonal or diagonal, respectively. Performing the Fourier transform vastly accelerates full quantum calculations and will allow us to calculate exact results below.\cite{kosloff1988time}

\subsection{\label{subsec:IAFSSH}Instantaneous Adiabatic Fewest Switches Surface Hopping (IA-FSSH)}
Fewest switches surface hopping formalism (FSSH) has emerged to be one of the most popular formalisms to describe non-adiabatic processes during the last thirty years. 
The nuclear degrees of freedom are treated as classical trajectories moving along the adiabatic potential energy surfaces and described by Newton's equation of motion. 
All non-adiabatic transitions are simulated by hopping process, which depend on the electronic amplitudes (which are propagated by the resulting time-dependent electronic Schr\"odinger equation and the non-adiabatic couplings).

As one moves to a time-dependent regime, as proposed by Gonz{\'a}lez et al,\cite{richter2011sharc, mai2018nonadiabatic} an intuitive way to extend the formalism is to replace the adiabatic basis $\ket{\phi^{ad}_{j}(\vec{R})}$, which parametrically depends on nuclear coordinate $\vec{R}$ in FSSH, by the instantaneous adiabatic electronic basis $\ket{\phi^{ad}_{j}(\vec{R}, t)}$, which parametrically depends on both nuclear coordinate $\vec{R}$ and time $t$.
\begin{eqnarray}
\hat{H}_{el}(\vec{R}, t)\ket{\phi^{ad}_{j}(\vec{R}, t)} = V^{el}_{j}(\vec{R}, t)\ket{\phi^{ad}_{j}(\vec{R}, t)}
\label{eq:four}
\end{eqnarray}
If we then expand the electronic wave function in this instantaneous adiabatic electronic basis,
\begin{eqnarray}
\ket{\Psi(\vec{R}, t)} \equiv \sum_{j}c_{j}(\vec{R}, t)\ket{\phi^{ad}_{j}(\vec{R}, t)}
\label{eq:five}
\end{eqnarray}
the electronic equation of motion becomes
\begin{eqnarray}
i\hbar \frac{\partial c_{j}(\vec{R}, t)}{\partial t} = H^{el}_{j}(\vec{R}, t)c_{j}(\vec{R}, t) - i\hbar \sum_{k}T_{jk}c_{k}(\vec{R}, t).
\label{eq:six}
\end{eqnarray}
Here, $T_{jk}$ is the time-derivative coupling matrix element defined as
\begin{eqnarray}
T_{jk}(t+dt_c/2) = \bracket{\phi_{j}^{ad}(\vec{R}, t)}{\frac{d \phi_{k}^{ad}(\vec{R}, t+dt_c)}{d t}}.
\label{eq:seven}
\end{eqnarray}
Note that there are two contributions to the time-derivative coupling matrix:
\begin{eqnarray}
\ket{\frac{d \phi_{k}^{ad}(\vec{R}, t)}{d t}}=\ket{\frac{\partial \phi_{k}^{ad}(\vec{R}, t)}{\partial R}} \frac{d\vec{R}}{dt} + \ket{\frac{\partial \phi_{k}^{ad}(\vec{R}, t)}{\partial t}}.
\label{eq:eight}
\end{eqnarray}
Here $\frac{d\vec{R}}{dt}$ is the nuclear velocity. The first term arises from the standard change of basis as induced by the nuclear velocity plus the time-independent non-adiabatic coupling. The second term is new and comes directly from the external field that changes in time.

According to Refs.~[\onlinecite{richter2011sharc}] and~[\onlinecite{mai2018nonadiabatic}], because a molecule can absorb energy from light, the IA-FSSH algorithm accepts all hops regardless of any energy restrictions: energy conservation and velocity rescaling are not enforced.\footnote{In principle, as the authors note, one should require that the energetic transition lie within the bandwidth of the applied EM field.}
\subsection{Floquet Theory}
Before introducing the F-FSSH formalism, let us briefly review Floquet theory as applied to solving a purely electronic TDSE.
For the problem with periodic Hamiltonian $\hat{H}(t) = \hat{H}(t+T_{0})$, the time-dependent electronic Schr\"odinger equation is
\begin{eqnarray}
i\hbar\frac{\partial}{\partial t}\ket{\Psi(t)} = \hat{H}(t)\ket{\Psi(t)}
\label{eq:TDSE}
\end{eqnarray}
and the initial wavefunction is 
\begin{eqnarray}
\ket{\Psi(t=0)} = \sum_{\mu}c_{\mu}(t=0)\ket{\mu} 
\label{eq:initwf}
\end{eqnarray}
Here, $\ket{\mu}$ is a complete, time-independent basis.

Floquet theory expands the solution to Eq.~(\ref{eq:TDSE}) in the following form
\begin{eqnarray}
\ket{\Psi(t)} = \sum_{m\mu}\tilde{c}_{m\mu}(t)\exp{(im\omega t)}\ket{\mu}
\label{eq:diab}
\end{eqnarray}
or, for short,
\begin{eqnarray}
\ket{\Psi(t)} =\sum_{m\mu}\tilde{c}_{m\mu}(t)\ket{\Xi^{m\mu}(t)}
\end{eqnarray}
where,
\begin{eqnarray}
\ket{\Xi^{m\mu}(t)} \equiv \exp{(im\omega t)}\ket{\mu}
\label{eq:fbas}
\end{eqnarray}

Here, $\ket{\Xi^{m\mu}(t)}$ is the Floquet basis, $\{\ket{\mu}\}$ is a set of electronic basis function and \textit{m} formally runs from $-\infty$ to $\infty$ (though in practice, a cutoff is applied).
If we plug Eq.~(\ref{eq:diab}) into the TDSE (Eq.~(\ref{eq:TDSE})), we find
\begin{eqnarray}
\sum_{m\mu}i\hbar\frac{\partial\tilde{c}_{m\mu}(t)}{\partial t}\ket{\Xi^{m\mu}(t)} + \sum_{m\mu}\tilde{c}_{m\mu}(t)i\hbar\frac{\partial}{\partial t}\ket{\Xi^{m\mu}(t)}
=\sum_{m\mu}\tilde{c}_{m\mu}(t)\hat{H}(t)\ket{\Xi^{m\mu}(t)} \nonumber
\end{eqnarray}
\begin{eqnarray}
\sum_{m\mu}i\hbar\frac{\partial\tilde{c}_{m\mu}(t)}{\partial t}\ket{\Xi^{m\mu}(t)}
=\sum_{m\mu}\tilde{c}_{m\mu}(t)\Bigl(\hat{H}(t) - i\hbar\frac{\partial}{\partial t}\Bigr)\ket{\Xi^{m\mu}(t)}
\label{eq:eleven}
\end{eqnarray}
We define the Floquet Hamiltonian as
\begin{eqnarray}
\hat{H}_{F}(t) \equiv \hat{H}(t) - i\hbar\frac{\partial}{\partial t}
\label{eq:twelve}
\end{eqnarray}
so that Eq.~(\ref{eq:twelve}) becomes
\begin{eqnarray}
\sum_{m\mu}i\hbar\frac{\partial\tilde{c}_{m\mu}(t)}{\partial t}\exp{(im\omega t)}\ket{\mu}
=\sum_{m\mu}\tilde{c}_{m\mu}(t)\hat{H}_{F}(t)\exp{(im\omega t)}\ket{\mu}.
\end{eqnarray}

At this point, we use the fact that $\hat{H}(t)$ is periodic in time so that we can expand an arbitrary matrix element of $\hat{H}_{F}(t)$ as a Fourier sum
\begin{eqnarray}
\bra{\nu}\hat{H}_{F}(t)\ket{\Xi^{m\mu}} = \sum_{n}\hat{\cal H}^{F}_{(n\nu)(m\mu)} \exp{[in\omega t]}
\label{eq:FHEXP}
\end{eqnarray}
$\hat{\cal H}^{F}_{(n\nu)(m\mu)}$ is the Fourier transform of the Floquet Hamitonian $\hat{H}_{F}$ which can be easily computed by inversion
\begin{eqnarray}
[\hat{\cal H}^{F}]_{(n\nu)(m\mu)} &=& \frac{1}{T_{0}}\int_{0}^{T_{0}}dt\bra{\nu}\hat{H}_{F}\ket{\Xi^{m\mu}}\exp[-in\omega t]
\nonumber \\
&=&\frac{1}{T_{0}}\int_{0}^{T_{0}}dt\bra{\nu}\hat{H}(t)\ket{\mu}\exp[-i(n-m)\omega t] + \delta_{\mu\nu}\delta_{m n}n\hbar\omega
\label{eq:sixteen}
\end{eqnarray}
where we have used the identity Eq.~(\ref{eq:fbas}).
Note that $\hat{\cal H}^{F}$ is Hermitian because $\hat{H}(t)$ is Hermitian.
Now if we multiply both sides of Eq.~(\ref{eq:eleven}) by $\bra{\nu}$, and gather all terms proportional to $\exp(in\omega t)$, we find
\begin{eqnarray}
i\hbar\frac{\partial\tilde{c}_{n\nu}(t)}{\partial t} = \sum_{m\mu}\tilde{c}_{m\mu}(t)\hat{\cal H}^{F}_{(n\nu)(m\mu)}
\label{eq:FTDSE}
\end{eqnarray}
Eq.~(\ref{eq:FTDSE}) can clearly be solved with the exponential operator because the entire $\hat{\cal H}_{F}$ is time-independent.
\begin{eqnarray}
\tilde{c}_{n\nu} = \sum_{m\mu}\Bigl[e^{-i\hat{\cal H}_{F}t/\hbar}\Bigr]_{(n\nu)(m\mu)}\tilde{c}_{m\mu}
\end{eqnarray}
In this sense, it is natural to focus on the eigenstates of the Floquet Hamiltonian, defined by
\begin{eqnarray}
\hat{\cal H}_{F}\ket{G^{J}}=\varepsilon_{J}\ket{G^{J}}.
\label{eq:TIDSE}
\end{eqnarray}

Clearly, in a Floquet eigenbasis, propagating the TDSE is simple. Let
\begin{eqnarray}
\ket{\Phi^{J}(t)} = \sum_{m\mu}G^{J}_{m\mu}\exp{(im\omega t)}\ket{\mu},
\end{eqnarray}
so that
\begin{eqnarray}
\hat{H}_{F}\ket{\Phi^{J}(t)}&=&\sum_{m\mu}(\sum_{n\nu}(\hat{\cal H}_{F})_{(m\mu)(n\nu)}G^{J}_{n\nu})\exp{(im\omega t)}\ket{\mu}\nonumber\\
&=&\varepsilon_{J}\sum_{m\mu}G^{J}_{m\mu}\exp{(im\omega t)}\ket{\mu}=\varepsilon_{J}\ket{\Phi^{J}(t)}.
\label{eq:twenty-two}
\end{eqnarray}
Then if we expand:
\begin{eqnarray}
\ket{\Psi(t)} \equiv \sum_{J}\tilde{c}_{J}(t)\ket{\Phi^{J}(t)},
\label{eq:expanad}
\end{eqnarray}
Eq.~(\ref{eq:TDSE}) becomes
\begin{eqnarray}
\sum_{J}i\hbar\frac{\partial\tilde{c}_{J}(t)}{\partial t}\ket{\Phi^{J}(t)}
=\sum_{J}\tilde{c}_{J}(t)\varepsilon_{J}\ket{\Phi^{J}(t)},
\label{eq:ten3}
\end{eqnarray}
whose solution is simply
\begin{eqnarray}
\tilde{c}_{J}(t)=\tilde{c}_{J}(t=0)\exp(-i\varepsilon_{J}t/\hbar)
\end{eqnarray}
Solving Eq.~({\ref{eq:TIDSE}}) is a simple mean of propagating the TDESE in a periodic field.

\subsection{Wavepacket dynamics with Floquet states}
Now, let us review how Floquet theory is applied to wavepacket dynamics with nuclear motion involved. The TDSE for the total nuclear-electronic wavefunction with nuclear coordinate $\vec{R}$ is:
\begin{eqnarray}
i\hbar\frac{\partial}{\partial t}\ket{\Psi(t)} = \hat{H}_{\text{tot}}(t)\ket{\Psi(t)} = \hat{\mathbb{T}}_{R} + \hat{H}_{el}(t)\ket{\Psi(t)}
\label{eq:TDSE2}
\end{eqnarray}
The initial electronic wavefunction at nuclear position $\vec{R}$ is
\begin{eqnarray}
\bracket{\vec{R}}{\Psi(t=0)} = \sum_{\mu}c_{\mu}(\vec{R}, t=0)\ket{\mu} 
\label{eq:initwff}
\end{eqnarray}
According to Floquet theory, we express the electronic wavefunction in Eq.~({\ref{eq:TDSE2}}) in the following form:
\begin{eqnarray}
\bracket{\vec{R}}{\Psi(t)} = \sum_{J}\tilde{c}_{J}(\vec{R}, t) \ket{\Phi^{J}(\vec{R}, t)}.
\label{eq:threeb}
\end{eqnarray}
where for now $J$ can be any electronic basis.

By plugging Eq.~(\ref{eq:threeb}) into the TDSE (Eq.~(\ref{eq:TDSE2})),
we obtain
\begin{eqnarray}
i\hbar \displaystyle \sum_{J} \frac{\partial \tilde{c}_{J}(\vec{R}, t)}{\partial t}\ket{\Phi^{J}(\vec{R}, t)}
+ i\hbar \displaystyle \sum_{J} \tilde{c}_{J}(\vec{R}, t)\frac{\partial \ket{\Phi^{J}(\vec{R}, t)}}{\partial t}
&=&(\hat{\mathbb{T}}_{\vec{R}} + \hat{H}_{el}) \ket{\Psi(\vec{R}, t)}
\nonumber\\
\Rightarrow i\hbar \displaystyle \sum_{J} \frac{\partial \tilde{c}_{J}(\vec{R}, t)}{\partial t}\ket{\Phi^{J}(\vec{R}, t)}
=\hat{\mathbb{T}}_{\vec{R}} \Bigl(\displaystyle \sum_{J} \tilde{c}_{J}(\vec{R}, t)\ket{\Phi^{J}(\vec{R}, t)}\Bigr) &+& \sum_{J}\tilde{c}_{J}(\vec{R}, t)\hat{H}_{F}^{el}(t) \ket{\Phi^{J}(\vec{R}, t)}
\label{eq:expanf}
\end{eqnarray}
Again, we have defined the electronic Floquet Hamiltonian as
\begin{eqnarray}
\hat{H}_{F}^{el}(t) \equiv \hat{H}_{el}(t) - i\hbar\frac{\partial}{\partial t}
\label{eq:twentysix}
\end{eqnarray}
The nuclear TDSE in Eq.~({\ref{eq:expanf}}) can be solved either in a diabatic electronic basis (where $J=(m\mu)$) or an adiabatic basis (where J labels a Floquet state).
\subsubsection{Wavepacket dynamics in a Diabatic Floquet representation}
Assuming a diabatic representation,
\begin{eqnarray}
\ket{\Phi^{J}(t)} \equiv \ket{\Xi^{m\mu}(t)} = \exp{(im\omega t)}\ket{\mu},
\end{eqnarray}
we can solve the TDSE in Eq.~({\ref{eq:expanf}}) by expanding the Floquet Hamiltonian (Eq.~(\ref{eq:twentysix})) as a function of time as in Eq.~(\ref{eq:FHEXP}) and comparing only terms with the same Fourier mode $\exp(in\omega t)$ and electronic state $\ket{\nu}$. We will obtain an equation similar to Eq.~(\ref{eq:FTDSE})
\begin{eqnarray}
i\hbar \displaystyle \frac{\partial \tilde{c}_{n\nu}(\vec{R}, t)}{\partial t}
=\hat{\mathbb{T}}_{\vec{R}} \tilde{c}_{n\nu}(\vec{R}, t) &+& \sum_{m\mu}\hat{\cal H}^{F}_{(n\nu)(m\mu)}\tilde{c}_{m\mu}(\vec{R}, t)
\label{eq:wpdf}
\end{eqnarray}
Eq.~(\ref{eq:wpdf}) is effectively a time-independent Schr\"odinger equation and can be easily solved.

\subsubsection{Wavepacket dynamics under adiabatic Floquet representation}
Assuming an adiabatic representation, the relevant Floquet basis is 
\begin{eqnarray}
\ket{\Phi^{J}(\vec{R}, t)} = \sum_{\mu m}G^{J}_{m\mu}(\vec{R})\exp{(im\omega t)}\ket{\mu},
\end{eqnarray}

By plugging the adiabatic Floquet basis into Eq.~({\ref{eq:expanf}}) and using the fact that the eigenvalues of $\hat{H}_{F}(t)$ are time-independent (for a periodic electronic Hamiltonian, see Eq.~({\ref{eq:twenty-two}}))
\begin{eqnarray}
\hat{H}^{el}_{F}\ket{\Phi^{J}(\vec{R}, t)} = \varepsilon(\vec{R}) \ket{\Phi^{J}(\vec{R}, t)},
\label{eq:eltidse}
\end{eqnarray}
we find
\begin{eqnarray}
i\hbar \displaystyle\sum_{J} \frac{\partial \tilde{c}_{J}(\vec{R}, t)}{\partial t}&\ket{\Phi^{J}(\vec{R}, t)}
=\displaystyle \sum_{J}\varepsilon_{J} \tilde{c}_{J}(\vec{R}, t) \ket{\Phi^{J}(\vec{R}, t)} 
\nonumber
\\
&+ \hat{\mathbb{T}}_{\vec{R}} \Bigl(\displaystyle \sum_{J} \tilde{c}_{J}(\vec{R}, t)\ket{\Phi^{J}(\vec{R}, t)}\Bigr).
\label{eq:twentynine}
\end{eqnarray}

If we apply $\bra{\nu} \otimes$ and compare only those terms with the same Fourier mode, $\exp(in\omega t)$, we find:
\begin{eqnarray}
&&\sum_{J}i\hbar \frac{\partial \tilde{c}_{J}(\vec{R}, t)}{\partial t}G^{J}_{n\nu}(\vec{R})
\nonumber \\
=&&\sum_{J}\varepsilon_{J}\tilde{c}_{J}(\vec{R}, t)G^{J}_{n\nu}(\vec{R}) 
+ \sum_{J}\hat{\mathbb{T}}_{\vec{R}}\bigl(\tilde{c}_{J}(\vec{R}, t)G_{n\nu}^{J}(\vec{R})\bigr)
\nonumber \\
=&& \sum_{J}\varepsilon_{J}\tilde{c}_{J}(\vec{R}, t)G^{J}_{n\nu}(\vec{R}) 
-\sum_{J}\frac{\hbar^2}{2M}\bigl(\nabla_{\vec{R}}^2\ \tilde{c}_{J}(\vec{R}, t)\bigr)\ G^{J}_{n\nu}(\vec{R}) 
\nonumber \\
&&- \sum_{J}\frac{\hbar^2}{M}\bigl(\nabla_{\vec{R}}\tilde{c}_{J}(\vec{R}, t)\bigr)\bigl(\nabla_{\vec{R}}G^{J}_{n\nu}(\vec{R})\bigr)
\nonumber \\
&&- \sum_{J}\frac{\hbar^2}{2M}\tilde{c}_{J}(\vec{R}, t)\bigl(\nabla_{\vec{R}}^2\ G^{J}_{n\nu}(\vec{R})\bigr)
\nonumber \\
\end{eqnarray}
After applying $\displaystyle \sum_{n\nu}G^{K *}_{n\nu}\otimes$ on both sides, and using the fact that
\begin{eqnarray}
\sum_{n\nu}G^{K*}_{n\nu}G^{J}_{n\nu} =  \delta_{JK},
\label{eq:thirtyfour}
\end{eqnarray}
we obtain
\begin{eqnarray}
i\hbar &&\frac{\partial \tilde{c}_{K}(\vec{R}, t)}{\partial t} = \varepsilon_{K}(\vec{R})\tilde{c}_{K}(\vec{R}, t)-\frac{\hbar^2}{2M}\nabla_{\vec{R}}^2\tilde{c}_{K}(\vec{R}, t) 
\nonumber
\\
&&- \frac{\hbar^2}{M}\sum_{J}\nabla_{\vec{R}}\tilde{c}_{J}(\vec{R}, t) \bracket{G^{K}(\vec{R})}{\nabla_{\vec{R}}\ G^{J}(\vec{R})} 
\nonumber
\\
&&- \frac{\hbar^2}{2M}\sum_{J}\tilde{c}_{J}(\vec{R}, t)\bracket{G^{K}(\vec{R})}{\nabla_{\vec{R}}^2\ G^{J}(\vec{R})}.
\label{eq:thirteen}
\end{eqnarray}
Again, Eq.~({\ref{eq:thirteen}}) appear to be entirely time-independent and can be easily solved.
Note that the probability to measure electronic state $\ket{\nu}$ is
\begin{eqnarray}
\bigl|\bracket{\nu}{\Psi(t)}\bigr|^2&=& \int d\vec{R}\Bigl|\sum_{J}\tilde{c}_{J}(\vec{R}, t)\sum_{m}G_{m\nu}^{J}(\vec{R})\exp(im\omega t)\Bigr|^2
\nonumber \\\label{eq:coherentsum}
&=&\int d\vec{R}\sum_{J}\Bigl|\tilde{c}_{J}(\vec{R}, t)\sum_{m}G_{m\nu}^{J}\Bigr|^2 
\\\nonumber &&+ \int d\vec{R}\sum_{J, K}\tilde{c}_{J}(\vec{R}, t)\tilde{c}^{*}_{K}(\vec{R}, t)\sum_{m\neq n}G_{n\nu}^{J}(\vec{R})(G_{m\nu}^{K}(\vec{R}))^{*}\exp[i(n-m)\omega t]
\end{eqnarray}
In Eq.~({\ref{eq:coherentsum}}), we find two terms that must be added together to find the total probability on state $\ket{\nu}$.

In the limit of no coupling (when adiabats and diabats are identical), $G^{J}_{m\nu} = \delta_{J, (m\nu)}$, and
\begin{eqnarray}
\bigl|\bracket{\nu}{\Psi(t)}\bigr|^2=\int d\vec{R}\sum_{m}\Bigl|\tilde{c}_{m\nu}(\vec{R}, t)\Bigr|^2 + \int d\vec{R}\sum_{m\neq n}\tilde{c}_{n\nu}(\vec{R}, t)\tilde{c}^{*}_{m\nu}(\vec{R}, t)\exp[i(n-m)\omega t]
\label{eq: coherentsumnocoupling}
\end{eqnarray}
\subsection{F-FSSH algorithm}
At last, we can present the F-FSSH algorithm. In general, surface hopping is valid only in an adiabatic representation.\cite{subotnik2016understanding} The nuclear motion in F-FSSH is described by Newton's equations of motion
\begin{eqnarray}
\dot{\vec{R}}=&&\vec{v} 
\nonumber
\\
\dot{\vec{v}} =&& -\frac{\nabla_{\vec{R}}\varepsilon_{J}(\vec{R})}{M}
\nonumber
\end{eqnarray}
where, $\varepsilon_{J}(\vec{R})$ is the adiabatic quasi-energy of Floquet state $J$, see Eq.~({\ref{eq:eltidse}}). After dropping the second derivative coupling term in Eq.~({\ref{eq:thirteen}}), the corresponding equations of motion for the electronic degrees of freedom are
\begin{eqnarray}
i\hbar \frac{\partial{\tilde{c}_{J}(\vec{R}, t)}}{\partial t} = &&\varepsilon_{J}(\vec{R})\tilde{c}_{J}(\vec{R}, t) \\ \nonumber
&&- i\hbar \sum_{K}\vec{v}\cdot \vec{d}_{JK} \tilde{c}_{K}(\vec{R}, t)  \\
\nonumber
=&& \varepsilon_{J}(\vec{R})\tilde{c}_{J}(\vec{R}, t) - i\hbar \sum_{K}T_{JK} \tilde{c}_{K}(\vec{R}, t).
\label{eq:seventeen}
\end{eqnarray}
Here, $\vec{d}_{JK}$ is the derivative coupling $\bracket{G^{K}(\vec{R})}{\nabla_{\vec{R}}\ G^{J}(\vec{R})}$ between Floquet state $J$ and $K$. $T$ is the time-derivative coupling matrix (see Eq.~(\ref{eq:seven})).
According to F-FSSH, the hopping probability from active Floquet state $J$ to state $K$ is
\begin{eqnarray}
g_{JK} = \frac{-2 \text{Re}(\tilde{c}_{J}\tilde{c}^{*}_{K}T_{KJ})}{|\tilde{c}_{J}|^2}dt
\label{eq:fifteen}
\end{eqnarray}
If $g_{JK}$ is less than $0$ for any $K$, we set $g_{JK}=0$.

After each successful hop, the velocity is adjusted to conserve the total Floquet quasi-energy. Unlike IA-FSSH, frustrated hops are not allowed.

\subsection{Nuances of the F-FSSH algorithm}
For exact wavepacket calculations in a Floquet basis, the exact total probability on a given electronic state $\nu$ is calculated by the coherent sum in Eq.~(\ref{eq: coherentsumnocoupling}). Thus, for surface hopping to match exact wavepacket dynamics, we must evaluate both the diagonal and interference terms in Eq.~(\ref{eq: coherentsumnocoupling}). The diagonal contribution is simple: 
\begin{eqnarray}
\text{Prob}^{\text{pop}}_{\nu} &=& \frac{\sum_{m}N^{\text{traj}}_{m\nu}}{N_{\text{tot}}}.
\label{eq:pop}
\end{eqnarray}
Here, $N_{\text{tot}}$ is the number of all independent trajectories, and $N^{\text{traj}}_{m\nu}$ is the number of the trajectories ending up asymptotically on a Floquet state with indices $m\nu$.

For the interference term, we will calculate two different variants. The first option is: 
\begin{eqnarray}
\text{Prob}^{\text{interference \#1}}_{\nu} = \sum_{n\neq m}\frac{\sum_{r=1}^{N_{n\nu}^{\text{traj}}}\sum_{s=1}^{N_{m\nu}^{\text{traj}}}\tilde{c}^{r}_{n\nu}(\tilde{c}^{s}_{m\nu})^{*}\exp(i(n-m)\omega t)}{{N^{\text{traj}}_{n\nu}\times N^{\text{traj}}_{m\nu}}}.
\label{eq:FFSSH_interf}
\end{eqnarray}
Here, $\tilde{c}^{r}_{n\nu}$ is the electronic amplitude on Floquet state $(n\nu)$ at the end of $r$th trajectory.

The second option is:
\begin{eqnarray}
\text{Prob}^{\text{interferece \#2}}_{\nu} =\sum_{n\neq m}\frac{\sum_{r=1}^{N_{n\nu}^{\text{traj}}}\sum_{s=1}^{N_{m\nu}^{\text{traj}}}\tilde{c}^{r}_{n\nu}(\tilde{c}^{s}_{m\nu})^{*}\exp(i(n-m)\omega t)\exp{\Bigl(-\frac{(P^{r}_{n\nu} - P^{s}_{m\nu})^2\sigma^2}{4\hbar^2}\Bigr)}}{{N^{\text{traj}}_{n\nu}\times N^{\text{traj}}_{m\nu}}}.
\label{eq:FFSSH_damp}
\end{eqnarray}
In Eq.~(\ref{eq:FFSSH_damp}), $P^{r}_{n\nu}$ is the instantaneous momentum at the end of the $r$th trajectory that ends up on Floquet state $(n\nu)$. The width $\sigma$ represents the width of the nuclear wavepacket that must be known a priori --- it should presumably be simply the initial width of the wavepacket in the full quantum calculation. 

In Eqs.~(\ref{eq:FFSSH_interf}) and~(\ref{eq:FFSSH_damp}), we have attempted to calculate the interference term in Eq.~(\ref{eq: coherentsumnocoupling}), by using the electronic amplitude ($\tilde{c}_{J}(t)$) to give us phase information about wavepackets propagating on state $J$, in keeping with the density matrix interpretation of FSSH as given in Ref.~{[\onlinecite{landry2012recover}]} (See section \textrm{III} in Ref.~{[\onlinecite{landry2012recover}]}). We introduce the damping factor $\exp{\Bigl(-\frac{(P^{r}_{n\nu} - P^{s}_{m\nu})^2\sigma^2}{4\hbar^2}\Bigr)}$ in Eq.~(\ref{eq:FFSSH_damp}) because whereas the coherence between adiabat $\ket{0}$ and $\ket{1}$ is eventually destroyed within standard FSSH because of the force difference $\Delta F = |F_{0}-F_{1}|$, there is clearly a more complicated physical basis for decoherence within F-FSSH. After all, for a time-dependent problem with two states, even though exact Floquet theory may propagate infinitely many states, there are actually only two unique forces at any given point in space. Thus, the standard FSSH decoherence approaches (based on $\Delta F$) are not applicable.
Future work will be necessary to identify the correct decoherence term, ideally by comparing F-FSSH with the QCLE.\cite{donoso1998simulation, kapral1999mixed, subotnik2013can, kapral2016surface} For now, the damping factor in Eq.~(\ref{eq:FFSSH_damp}) comes simply from the overlap of two frozen gaussians with different momenta. All of the independent trajectories are propagated for a certain amount of time $t$, which is long enough for all trajectories to pass through the interaction zone.

\section{\label{sec:3}Simulation details}
In this paper, for simplicity, we will focus on one dimensional models and specifically time-periodic variants of the famous Tully model problems.\cite{tully1990molecular}
\subsection{Model problems\label{sec:3.1}}
Tully's simple avoided crossing model problem is modified in the diabatic representation as
\begin{eqnarray}
H^{el}_{00}(R) &=& A[1-\exp(-B\times R)], \quad  R>0,\nonumber \\
\label{eq:T1st}
H^{el}_{00}(R) &=& - A[1-\exp(B\times R)], \quad  R<0, \\\nonumber
H^{el}_{11}(R) &=& -H^{el}_{00}(R), \\ \nonumber
H^{el}_{01}(R, t) &=& H^{el}_{10}(R, t) = C\exp(-D\times R^2)\cos\omega t= V(R)\cos\omega t.
\end{eqnarray}
The parameters are the same as the original paper, $A = 0.01$, $B = 1.6$, $C = 0.005$, $D = 1.0$ and we will test two different omega cases, $\omega = 0.008, 0.012$.

Tully's dual avoided crossing model problem is modified in a similar fashion
\begin{eqnarray}
H^{el}_{00}(R) &=& 0, \nonumber\\
\label{eq:T2nd}
H^{el}_{11}(R) &=& -A\exp(-B\times R^2) + E_{0}, \\
H^{el}_{01}(R, t) &=& H^{el}_{10}(R, t) = C\exp(-D\times R^2)\cos\omega t = V(R)\cos\omega t. \nonumber
\end{eqnarray}
The parameters are $A = 0.10$, $B = 0.28$, $E_0 = 0.05$, $C = 0.015$, $D = 0.06$ and $\omega = 0.02, 0.04$.

As discussed in Section~{\ref{sec:one}} and~\ref{subsec:IAFSSH}, IA-FSSH propagates trajectories along the instantaneous PESs. To better understand IA-FSSH, note that according to Eqs.~(\ref{eq:T1st}) and~(\ref{eq:T2nd}), in our model problems, the number of avoided crossings remain the same and they remain at the same position, but the strength of the instantaneous diabatic coupling oscillates.

\begin{figure*}
\includegraphics[width=\textwidth]{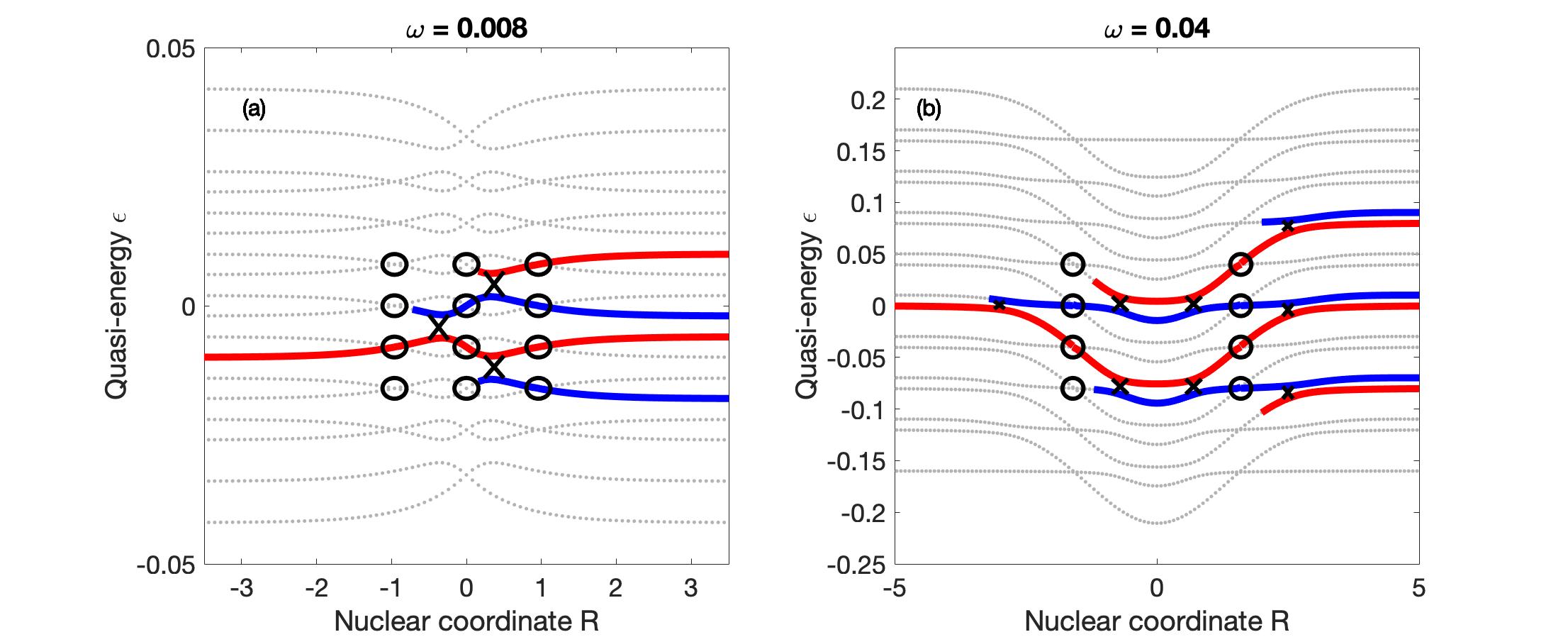}
\caption{\label{fig:Floquet_states} Floquet states and possible transmission Floquet channels for (a) the modified simple avoided crossing problem with $\omega = 0.008$ and (b) the modified dual avoided crossing problem with $\omega = 0.040$ (see Eqs.~(\ref{eq:T1st}) and~(\ref{eq:T2nd}). Both figures are truncated at $m=\pm 4$. The solid circles highlight trivial crossings and the solid arrows highlight real avoided crossings. The trajectories start from the left on the red state (diabat $\ket{0}$) and moving to the right. As particles move along the Floquet potential quasi-energy surface, trajectories may or may not hop between these Floquet states near avoided crossings and must hop at trivial crossings in order to remain on the same diabatic state.}
\end{figure*}

Unlike IA-FSSH, however, Floquet theory is more complicated. The potential quasi-energy surfaces generated are worth further discussion. The Floquet Hamiltonian $\hat{\cal H}_{F}^{el}$ after Fourier transform is shown below.
\setlength{\arraycolsep}{0.8pt}
\[\hat{\cal H}_{F}^{el}(R) = \left[
\begin{array}{c | c c | c c | c c | c}
\ddots & & & & & & & \\ \hline
 & H^{el}_{00}(R) + \hbar\omega &  &  & V(R)/2 &  & & \\ 
& & H^{el}_{11}(R) +\hbar \omega & V(R)/2 & &  & &\\ \hline
& & V(R)/2 & H^{el}_{00}(R) & & & V(R)/2 & \\ 
&  V(R)/2 & & & H^{el}_{11}(R) & V(R)/2 & &  \\ \hline
& & & & V(R)/2 & H^{el}_{00}(R) - \hbar\omega  &  & \\ 
& & & V(R)/2 & & & H^{el}_{11}(R) - \hbar\omega & \\ \hline
& & & & & &  & \ddots \\
\end{array} \right] \]
Here, $V(R)$ is the time-independent part of the coupling $H^{el}_{01}(R, t)$ and $H^{el}_{10}(R, t)$. The adiabatic potential quasi-energy surfaces are obtained by diagonalizing the electronic Floquet Hamiltonian $\hat{\cal H}_{F}^{el}$. As shown in Fig.~\ref{fig:Floquet_states}, relative to the original time-independent Tully models, F-FSSH can include many more than one avoided crossings, and light-induced trivial crossings (black circles) and avoided crossings (black crosses) are both possible. In particular, note that the original crossings in the time-independent models become a set of trivial crossings in Fig.~{\ref{fig:Floquet_states}}
\subsection{Exact Calculation}
As a benchmark for our semiclassical calculations, we will perform exact dynamic simulations. A Gaussian wavepacket is initialized on diabat $\ket{0}$
\begin{eqnarray}
\ket{\Psi(R, t=0)} &&
\nonumber \\
&&= \sqrt[4]{\frac{1}{\pi\sigma^2}}\exp\Bigl(-\frac{(R-R_{0})^2}{2\sigma^2} + iP_{0}(R-R_{0})\Bigr)\ket{0}.
\label{eq:simudeone}
\end{eqnarray}
Here, $R_{0}, P_{0}$ are the initial position and momentum, and $\sigma$ is the width of the Gaussian which is chosen to be $\sigma = 20\hbar/P_{0}$. The wavepacket is propagated by the propagator in Eq.~(\ref{eq:three}) with the full, instantaneous Hamiltonian.

\subsection{Initial Conditions}
For the exact, IA-FSSH and F-FSSH calculations, we choose $R_{0} = -9.0$ which is far enough from $R=0$ such that there is never any coupling at the initial time.

For F-FSSH, Floquet theory generates a set of Floquet states with different Fourier modes from a single electronic state. Note that to the left of the coupling regime, the diabatic and adiabatic electronic states are identical.  The independent trajectories are initialized on the Floquet state generated by the diabat $\ket{0}$ with Fourier index $0$, i.e. the Floquet quasi-energy is the same as the energy of the diabat $\ket{0}$ at $R_0$.
\subsection{Truncation of $m$}
In principle, the Fourier series in Eq.~(\ref{eq:FHEXP}) should sum from $-\infty$ to $\infty$, i.e. the Floquet Hamiltonian $\hat{\cal H}_{F}$ should be infinitely large, which would be impossible to diagonalize. However, as a practical matter, we can truncate highly oscillating states. In this paper, for the Hamiltonian in Eqs.~(\ref{eq:T1st}) and~(\ref{eq:T2nd}), all off-diagonal Hamiltonian matrix elements are of the form $V(R)\cos(\omega t)$, such that only $m=\pm 1$ changes are possible. To calculate the quasi-energies accurately, we find that truncation at $m=\pm 4$ is usually converged (unless otherwise noted).
\subsection{Separation of Time Scale between Classical and Quantum Degrees of Freedom}
For both F-FSSH and IA-FSSH, nuclear motion is propagated with classical time-step $dt_c$ and the electronic propagation are propagated with a smaller quantum time-step $dt_q$, which is determined by\cite{jain2016efficient}
\begin{equation}\label{eq: dt_q}
\begin{aligned} 
dt'_q = \min \left(
\begin{array} {cc} 
dt_c \\
0.02 / \max[{\epsilon_{K}} - \bar{\epsilon}] \\
0.02 / \max[{T}]
\nonumber
\end{array}
\right),
\end{aligned}
\end{equation}
\begin{eqnarray}
dt_{q} = \frac{dt_{c}}{\text{nint}(dt_{c}/dt'_{q})}.
\end{eqnarray}
Here, $\max[T]$ represents the greatest element in the time-derivative coupling matrix, $\bar{\epsilon}$ is the average of all quasi-energies $\{\epsilon_{K}\}$ and $\text{nint}(x)$ is the smallest integer that is greater than $x$. We check for a hop at every $dt_q$ time step, but there should be at most one successful hop within a single $dt_c$.

\subsection{\label{sec:derivative_coupling}Evaluating the derivative coupling $d_{JK}$}
When two diabatic states of a molecule cross with each other, two scenarios are possible. First, the crossing can be meaningful with a finite diabatic coupling, leading to a finite probability of switching diabats. Second, when there is effectively no coupling between surfaces, the molecule must always remain on the same diabat as it leaves the crossing point. The non-adiabatic coupling is either undefined or, with machine error, approaches infinity. This second type of crossings has been called a trivial crossing,\cite{hammes1994proton, fabiano2008implementation, barbatti2010non, plasser2012surface, fernandez2012identification, nelson2013artifacts,  wang2014simple, meek2014evaluation, jain2016efficient, lee2019solving} and often leads to numerical instabilities (requiring very small time steps). Trivial crossings are common in F-FSSH, see circles in Fig.~{\ref{fig:Floquet_states}}.

In order to alleviate the problem of trivial crossings, an overlap matrix scheme has been proposed in Refs. [\onlinecite{meek2014evaluation}] and [\onlinecite{jain2016efficient}]. The basic idea is to calculate the time-averaged derivative coupling matrix by first calculating the overlap matrix $U_{JK}$ at different times and second evaluating the time-derivative coupling matrix $T_{JK}$, which is the logarithm of $U_{JK}$.
\begin{eqnarray}
v&&d_{JK} = T_{JK} = \frac{1}{dt_c}\log[U_{JK}]
\\
U_{JK} &&= \bracket{\Phi^{J}(R(t))}{\Phi^{K}(R(t+dt_c))}
\label{eq:twenty}
\end{eqnarray}
The matrix logarithm can be stably realized by a Schur decomposition.\cite{loring2014computing} Of course, for the logarithm to be real, the signs of eigenvectors ${\ket{\Phi_{J}(R)}}$ must be adjusted to guarantee that the overlap matrix $U_{JK}$ is a rotation matrix.

Unfortunately, one shortcoming of the overlap approach just mentioned is the fact that choosing the signs of each adiabatic state can be non-trivial. 
Then for example, all sign combinations are possible for $U$:
\[MU = \left[
\begin{array}{c c c c}
\pm 1 & & &\\
&  \pm 1 & &\\
& & \ddots & \\
& & & \pm 1\\
\end{array} \right] U\]
There are $2^N$ possible matrices for $U$, where $N$ is the size of the matrix $U$.
With this problem in mind, in a companion paper,\cite{zeyu2019robust} we show that choosing the sign of U to minimize $||\log(MU)||$ is a stable, practical and accurate ansatz. We believe this should definitively solve the trivial crossing sign problem and we will have further discussion on the more general form of this algorithm in the following publication.\cite{zeyu2019robust}
\subsection{Velocity Reversal}
Implementing velocity reversal after a frustrated hop is known to be crucial when calculating non-adiabatic rates (e.g. Marcus theory\cite{jain2015surface}). Velocity reversal improves the branching ratios in the surface hopping algorithm.\cite{jasper2003improved}
For the one-dimensional model problems below, we will reverse velocity if a frustrated hop between active state $\lambda$ and target state $\eta$ occurs and if
$v F_{\eta} < 0$ and $F_{\eta} F_{\lambda} < 0$.

\subsection{Obtaining the electronic amplitudes and momenta for calculating the interference terms in Eqs.~(\ref{eq:FFSSH_interf}) and~(\ref{eq:FFSSH_damp})}
For the scattering calculations below, we will record the electronic amplitude of the active Floquet state and the instantaneous momentum in memory at the end of each trajectory. Afterwards, one must calculate the average interference term according to Eqs.~(\ref{eq:FFSSH_interf}) or~(\ref{eq:FFSSH_damp}). A few nuances should now be explained.

First, for FSSH, the signs of the electronic amplitudes are always defined by using the overlap matrix scheme in Section.~{\ref{sec:derivative_coupling}}. However, in principle, each trajectory can choose its own sign for the adiabatic electronic states. Thus, in order to calculate the interference term in Eqs.~(\ref{eq:FFSSH_interf}) and~(\ref{eq:FFSSH_damp}), we must be sure that the electronic states (at the end of each dynamical trajectory) all have a consistent definition. For that purpose, we must record the sign changes along each trajectory, multiply all overlap matrices and obtain the total overlap matrix $U^{tot}$ from the beginning to end:
\begin{eqnarray}
U^{tot} = \prod_{t=0}^{\infty} U(t)
\end{eqnarray}
For simplicity, we will initialize all calculations with adiabats equal to diabats and we will insist that asymptotically (at the end of the trajectory), the adiabats are still equal to the diabats. Thus, if $U^{tot}_{jk}=-1$, we will change the sign of $\tilde{c}_{k}$ at the end of the trajectory.

Second, asymptotically, far away from the crossings, each Floquet state corresponds to a single diabatic electronic state $\ket{\nu}$ and so, when evaluating Eq.~(\ref{eq:FFSSH_interf}) and Eq.~(\ref{eq:FFSSH_damp}), for the probability to occupy diabatic state $\ket{\nu}$, we sum over only those trajectories ending with active Floquet state corresponding to electronic state $\ket{\nu}$.

Third, for the two electronic amplitudes ($\tilde{c}^{r}_{n\nu}$ and $\tilde{c}^{s}_{m\nu}$) and momenta ($P^{r}_{n\nu}$ and $P^{s}_{m\nu}$) in Eq.~(\ref{eq:FFSSH_interf}) and Eq.~(\ref{eq:FFSSH_damp}), note that these sets correspond to trajectories ending on states with different Fourier indices $m$ and $n$ but the same $\nu$: we average over all combinations of trajectories ending on different $m\nu$ and $n\nu$.
We never average amplitudes over different trajectories ending up on the same $(m\nu)$ state: for these matrix elements, we use only the active surface to estimate populations in a fashion consistent with the density matrix interpretation of surface hopping trajectories (Method \#3 in Ref. [\onlinecite{landry2012recover}]).
\subsection{Normalization}
If we sum the probabilties to occupy all diabatic states $j$, we should ideally recover unity. However, in practice, when we use F-FSSH, we find a problem. On the one hand, the diagonal contribution in Eq.~(\ref{eq:pop}) is 1. On the other hand, the contributions from the interference terms in Eqs.~(\ref{eq:FFSSH_interf}) and~(\ref{eq:FFSSH_damp}) may not be equal and opposite with each other, which results in incomplete cancellation, i.e. $\sum_{\nu} \text{Prob}_{\nu}^{\text{interference \#1}} \neq 0$ and $\sum_{\nu} \text{Prob}_{\nu}^{\text{interference \#2}} \neq 0$. Beyond fixing up the decoherence problems with F-FSSH, there is no simple solution to this problem, and we will simply normalize our probability results. In most cases below, the total probability is not far from $1$ ($<5\%$).  

\section{\label{sec:4}RESULTS}
We now present IA-FSSH, F-FSSH and exact results for the modified Tully model problems presented above.
\begin{figure*}
\includegraphics[width=1.0\textwidth]{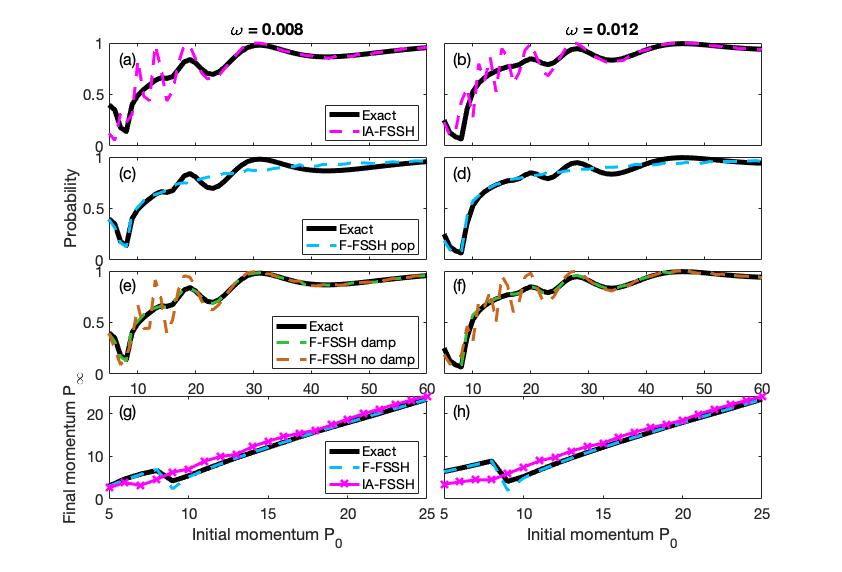}
\caption{\label{fig:T1st_instantad_F-FSSH} Transmission probabilities and the final momentum on diabat $\ket{0}$ for the modified simple avoided crossing problem with initial state on diabat $\ket{0}$ at $R = -9$. The figures (a), (c), (e), (g) on the left represent results with $\omega = 0.008$ and the four figures (b), (d), (f), (h) on the right side represent results with $\omega = 0.012$. (a), (b) plot the transmission probability on diabat $\ket{0}$ transmission channel for surface hopping with instantaneous adiabatic states (IA-FSSH) versus of exact dynamics. The results clearly show that IA-FSSH can recover the correct results only in the extremely high kinetic energy regime. (c), (d) plot according to Eq.~({\ref{eq:pop}}) for F-FSSH versus the exact results demonstrating the need to include interference at high momentum. (e), (f) plot exact results versus F-FSSH results as estimated by Eq.~({\ref{eq:FFSSH_interf}}) and Eq.~({\ref{eq:FFSSH_damp}}). The damping factor successfully corrects all results. (g), (h) plot the final momenta on diabat $\ket{0}$ transmission channel after passing through the interaction zone. The kinks at low momentum regime arise from opening up a new channel in Floquet space (see Fig.~{\ref{fig:Floquet_states}}) and the fact that both exact and F-FSSH results predict such a kink is a strong evidence that energy absorption/emission is treated properly with F-FSSH.}
\end{figure*}

\subsection{Modified Simple Avoided Crossing Problem}
\begin{figure*}
\includegraphics[width=1.0\textwidth]{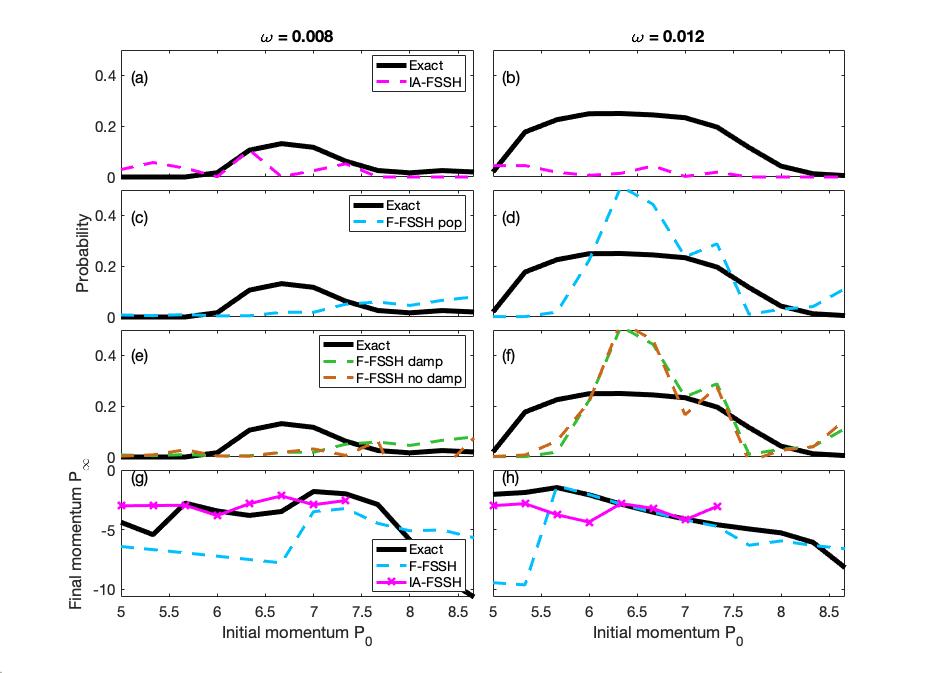}
\caption{\label{fig:Reflec1st_instantad_F-FSSH} Reflection probabilities and the final momentum on the diabat $\ket{1}$ for the modified simple avoided crossing problem. All figures represent the same physical quantities as in Fig.~{\ref{fig:T1st_instantad_F-FSSH}} except now we plot the reflection channel. Reflection onto diabat $\ket{1}$ does not occur for the time-independent simple avoided crossing problem. }
\end{figure*}
\subsubsection{IA-FSSH vs Exact}
The transmission probability results for IA-FSSH are shown in Figs.~{\ref{fig:T1st_instantad_F-FSSH}} (a), (b). The exact results increase smoothly as momentum gets larger in both $\omega$ cases ($\omega = 0.008, 0.012$). IA-FSSH accurately predicts the results for high momentum. However, for low momentum, the results from IA-FSSH oscillate rapidly and strongly while the exact result is smooth.

To test whether the energy absorption/emission is correctly predicted, we calculated the final average momentum. As shown in Figs.~{\ref{fig:T1st_instantad_F-FSSH}} (g), (h), the exact results show a kink at $P_0 = 9$. This kink comes from the fact that as the initial momentum increases, the wavepacket has just enough energy to move along diabat $\ket{0}$. This effect cannot be recovered by IA-FSSH, which predicts that the transmitted momentum increases steadily as the initial momentum grows.

Next, as shown in Fig.~{\ref{fig:Reflec1st_instantad_F-FSSH}}, according to exact dynamics, the time-dependent model problem yields a new reflection channel through diabatic state $\ket{1}$, in comparison to the original, time-independent model. Unfortunately, IA-FSSH can barely predict such a peak, even though the final momentum can be predicted approximately.
Obviously, the performance of IA-FSSH is mixed with respect to modified Tully model problem \#1.
\subsubsection{F-FSSH vs Exact}
Let us now discuss the performance of F-FSSH. The three possible approximations (Eqs.~(\ref{eq:pop}),~(\ref{eq:FFSSH_interf}) and~(\ref{eq:FFSSH_damp})) for the two $\omega$ cases are shown in Fig.~{\ref{fig:T1st_instantad_F-FSSH}} (c)-(f). The first approximation (Eq.~({\ref{eq:pop}}), labeled as "F-FSSH pop"), accurately predicts the results at lower momentum regime but cannot predict the oscillations at high momentum regime, as shown in Fig.~{\ref{fig:T1st_instantad_F-FSSH}} (c), (d). The second approximation (Eq.~({\ref{eq:FFSSH_interf}}), labeled as "F-FSSH no damp"), as shown in Fig.~{\ref{fig:T1st_instantad_F-FSSH}} (e), (f), is nearly the same as the IA-FSSH results and can only predict the exact results at high momentum regime. The third approximation (Eq.~({\ref{eq:FFSSH_damp}}), labeled as "F-FSSH damp") can predict the exact transmission probabilities for all regimes.
With regard to the final momentum, note that unlike IA-FSSH, F-FSSH does recover the correct kink in Fig.~{\ref{fig:T1st_instantad_F-FSSH}} (g), (h).

As for the new reflection channel, as shown in Fig.~{\ref{fig:Reflec1st_instantad_F-FSSH}} (c)-(f). F-FSSH can recover some reflection approximately in the case of $\omega = 0.012$, but not in the case of $\omega = 0.008$. In the case of $\omega = 0.008$, the reflected final average momentum can barely be predicted by F-FSSH, but the results are predicted qualitatively in the case of $\omega = 0.012$.

\subsection{Modified Dual Avoided Crossing Problem}
\begin{figure*}
\includegraphics[width=1.0\textwidth]{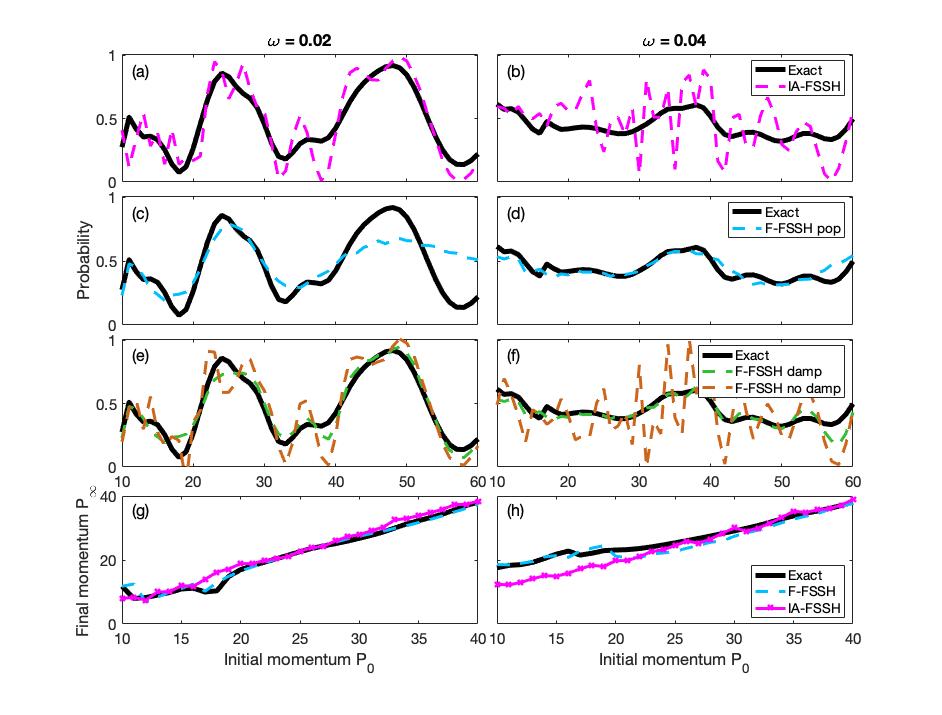}
\caption{\label{fig:trans_2nd_instantad_F-FSSH} Transmission probabilities and final momenta on diabat $\ket{1}$ for the modified dual avoided crossing problem. The four figures (a), (c), (e) and (g) on the left shows the transmission probabilities and final momentum results for $\omega = 0.02$ and the figures (b), (d), (f) and (h) on the right side shows the results for $\omega = 0.04$. Despite the fact that more states are involved in this model problem, F-FSSH with the interference term can still recover the exact results while IA-FSSH can hardly estimate the results in most regimes. The kinks in the final momentum are recovered by F-FSSH; see Figs.(g), (h)}
\end{figure*}
\subsubsection{IA-FSSH vs Exact}
\begin{figure*}
\includegraphics[width=1.0\textwidth]{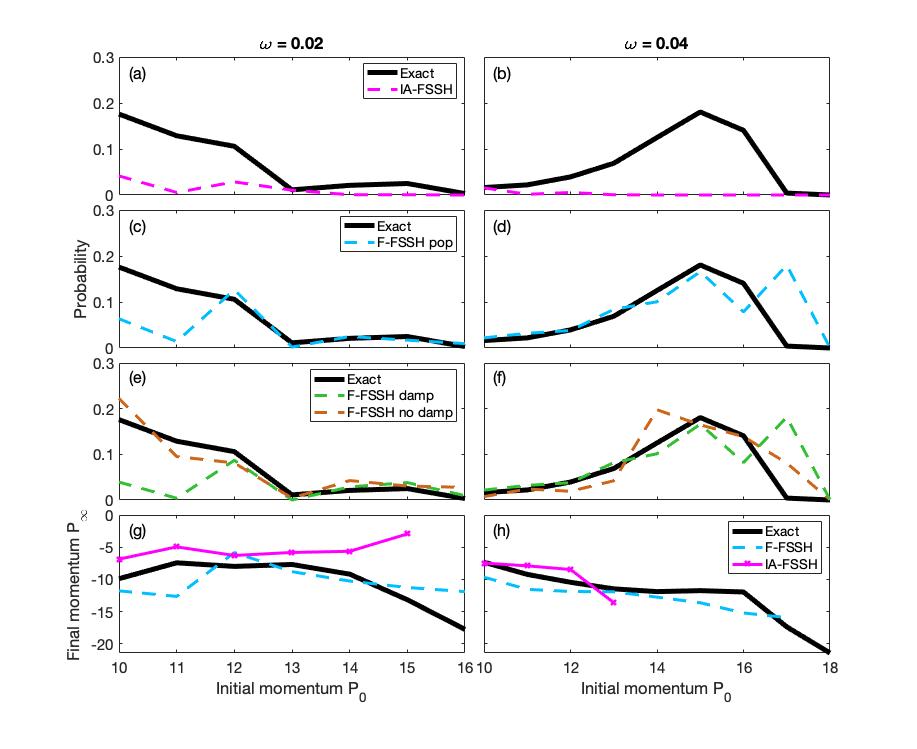}
\caption{\label{fig:reflecs_2nd_instantad_F-FSSH} Reflection probabilities on diabat $\ket{1}$ for modified dual avoided crossing problem. All figures represent the same quantities as Fig. \ref{fig:trans_2nd_instantad_F-FSSH} except now we plot the reflection channel. IA-FSSH predicts a very small reflection probability along this new channel while F-FSSH formalism recovers the correct trend. }
\end{figure*}

For the second modified Tully model problem, as shown in Fig.~{\ref{fig:trans_2nd_instantad_F-FSSH}} (a), (b), in the case of $\omega = 0.02$, the exact results oscillate strongly while in the case of $\omega = 0.04$, the exact results are almost flat. Here, we find an interesting nuance: for large frequencies of incoming light, many features of the potential energy surface become washed away and flat scattering probabilities arise. That being said, IA-FSSH misses this effect. IA-FSSH results oscillates strongly and rapidly in both cases and can predict the results only at high momentum regime in the case of $\omega = 0.02$.
As far as the final average momentum is concerned, IA-FSSH again cannot predict the kink as shown in Fig.~{\ref{fig:trans_2nd_instantad_F-FSSH}} (g), (h). 
As far as the reflection channel on diabatic state $\ket{1}$ is concerned, IA-FSSH can hardly predict the peak from the exact results, as shown in Figs.~{\ref{fig:reflecs_2nd_instantad_F-FSSH}} (a), (b). 
\subsubsection{F-FSSH vs Exact}
Unlike IA-FSSH, the F-FSSH is quite accurate for this model problem, even though multiple crossings are possible. The first approximation (Eq.~(\ref{eq:pop})) successfully predicts the exact results at low momentum regime. Similar to the previous case, as shown in Fig.~{\ref{fig:trans_2nd_instantad_F-FSSH}} (e), (f), the second approximation (Eq.~(\ref{eq:FFSSH_interf})) yields results that are nearly the same as the results of IA-FSSH. Finally, the third approximation (Eq.~(\ref{eq:FFSSH_damp}))  predicts excellent results (matching the exact results) in all regimes.
In Fig.~{\ref{fig:trans_2nd_instantad_F-FSSH}} (g), (h), the final average momentum and the kinks as predicted by F-FSSH agreement are also in close with the exact results. Moreover, for the reflection channel, as shown in Fig.~{\ref{fig:reflecs_2nd_instantad_F-FSSH}} (c)-(f), the F-FSSH formalism gives a pretty good estimate of the reflection probabilities (and the final average momenta is qualitatively correct).

Obviously, by reducing a time-dependent problem to a time-independent non-adiabatic problem, where standard FSSH applies and energy conservation can be enforced, F-FSSH becomes far more accurate than IA-FSSH.

\section{Discussion: Convergence to time-independent simple avoided crossing results}
While the results above strongly suggests that F-FSSH is promising (more so than IA-FSSH) for modeling non-adiabatic dynamics under illumination, one key point that has not yet been addressed is the $\omega \rightarrow 0$ limit.\cite{hone1997time}
As the frequency $\omega$ in Eq.~(\ref{eq:T1st}) and Eq.~(\ref{eq:T2nd}) approaches zero, all of our modified model Hamiltonians become the original, time-independent Tully model Hamiltonians and one must wonder: Will F-FSSH reduce to standard FSSH? 
The answer is not clear because although Floquet states will approach one another, they will always remain separate, each with their own individual dynamics.  In this limit, there will be many Floquet states involved and the role of interference term is likely to grow. Does F-FSSH agree with FSSH here?

\begin{figure*}
\includegraphics[width=\textwidth]{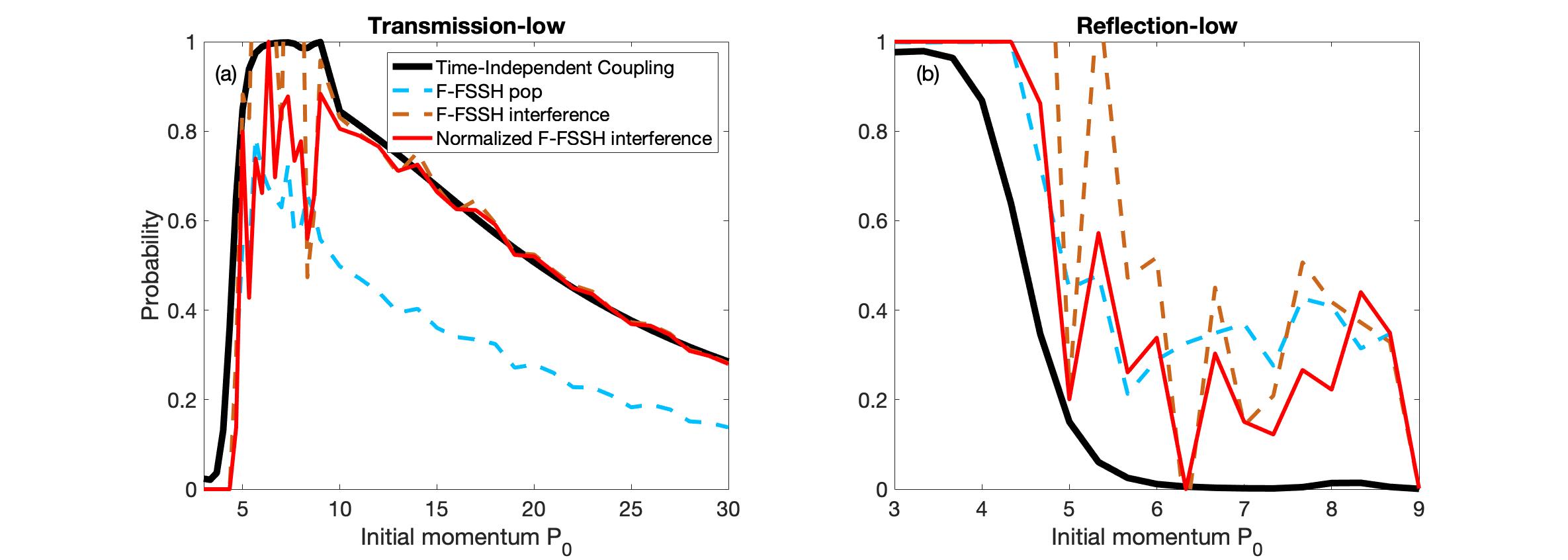}
\caption{\label{fig:lowomegaresult}Comparison between time-independent coupling simple avoided crossing model results and the F-FSSH results with $\omega = 0.00001$. Both the transmission (a) and reflection (b) probabilities at lower channels are shown above. The black line represents results for the time-independent Tully's simple avoided model problem. F-FSSH results for $\omega = 0.00001$ show convergence to the time-independent results only in the high momentum regime while including the interference term. In the low momentum regime, the interference term is highly oscillatory and cannot converge to the time-independent results even after normalization.}
\end{figure*}

To test this limit, we have run simulations. As shown in Fig.~{\ref{fig:lowomegaresult}} (a), at high momentum ($P_{0} > 10$), where there are only two transmission channels, the F-FSSH algorithm does reduce to time-independent results if we include the interference term. At low momentum, however, the interference term becomes highly oscillatory and F-FSSH no longer matches FSSH. As also shown in Fig.~{\ref{fig:lowomegaresult}} (b), the interference term is unstable and F-FSSH cannot predict the reflection probability on the reflection channel along $\ket{0}$.
As mentioned above, the coherence problem is not simple, since we must evaluate the coherence between Floquet states generated from the same electronic state and along which the forces are exactly same, so that previous work on decoherence with surface hopping is not obviously relevant.\cite{schwartz1994aqueous, bittner1995quantum, schwartz1996quantum, prezhdo1997evaluation, prezhdo1997mean, fang1999improvement, fang1999comparison, volobuev2000continuous, hack2001electronically, wong2002dissipative, wong2002solvent, horenko2002quantum, jasper2005electronic, bedard2005mean, subotnik2011decoherence, subotnik2011new, landry2012recover}
A new understanding of coherence and decoherence in the content of F-FSSH will be necessary and it would appear fruitful at this time to turn to the QCLE guidance,\cite{kapral1999mixed, horenko2001theoretical} or perhaps exact factorization.\cite{abedi2010exact, abedi2012correlated, suzuki2015laser, min2017ab}

In the meantime, how are we to run F-FSSH dynamics if $\omega$ changes and goes from finite to zero?
In Schmidt et al's original F-FSSH paper, \cite{fiedlschuster2016floquet} when dealing with light pulse, the authors switched between Born-Oppenheimer states and Floquet states for time intervals without or with light pulses. In this slightly modified version of F-FSSH formalism, there was no additional cost for switching back and forth between algorithms since one can always change the couplings in the Floquet Hamiltonian when the light pulse approaches. For now, if $\omega$ ever vanishes, we believe this combination of F-FSSH and FSSH is optimal. In particular, when $V \leq \hbar \omega$, we suggest running F-FSSH and switching to standard FSSH when $V \gg \hbar \omega$.

\section{\label{sec:5}Conclusions}
In conclusion, we have compared two surface hopping formalisms for extending FSSH into time-dependent regimes. For our model problems, neither transmission probabilities nor reflection probabilities can be recovered by IA-FSSH. The formalism gives strongly oscillatory results in most cases, corresponding to the fact that the instantaneous adiabatic electronic states depend strongly on time and completely ignore the changes in state as induced by dressing with photons. 
The F-FSSH formalism looks more promising, as it recovers accurate transmission probabilities and approximate reflection probabilities on each electronic state. By evolving along a set of time-independent Floquet states, the algorithm yields smooth results. In addition, the interference terms between different wavepackets can be estimated using F-FSSH. Most importantly, F-FSSH can approximately estimate the reflection on the excited state, which is a new reflection channel, while there is hardly any reflection probability according to IA-FSSH. That being said, F-FSSH is only qualitatively correct and sometimes far from quantatively correct. It is clear that more work will be necessary if we wish to optimize how we calculate the interference term in Eqs.~(\ref{eq:FFSSH_interf}) and~(\ref{eq:FFSSH_damp}). We believe that it will be extremely useful to revisit F-FSSH in the context of the QCLE (where we can likely learn more about decoherence).\cite{donoso1998simulation, kapral1999mixed, subotnik2013can, kapral2016surface} This work is ongoing.

With regards to nuclear observables, F-FSSH can yield a good estimate of the resulting momentum in each channel of reflection and transmission, as compared to the exact results. The kink in the middle momentum regime in Figs.~{\ref{fig:T1st_instantad_F-FSSH}} (g), (h) and Figs.~{\ref{fig:trans_2nd_instantad_F-FSSH}} (g), (h), represents the situation that the transmission channel along diabat $\ket{0}$ dressed with $0$ photon (see Fig.~{\ref{fig:Floquet_states}} (a) the top red state) and the channel along diabatic state $\ket{1}$ dressed with 2 photons (see Fig.~{\ref{fig:Floquet_states}} (a) the top blue state) start to participate and this transmission can be recovered by F-FSSH while the IA-FSSH washes out this effect.

As far as the model problem studied in this paper (Eq.~(\ref{eq:T1st}) and~(\ref{eq:T2nd})), we must note that our results for the time-dependent versions of Tully's model problem do depend on where the trajectories start $R(t=0)$. In Figs.~\ref{fig:T1st_instantad_F-FSSH} -~\ref{fig:reflecs_2nd_instantad_F-FSSH}, we initialized all trajectories to start at $R=-9.0$ at time $0$. Obviously, because the diabatic coupling is time-dependent, the outcome of a scattering event will depend critically on when a trajectory reachs the effective coupling region, or equivalently, the initial phase of the time-dependent diabatic coupling. In Eqs.~(\ref{eq:T1st}) and~(\ref{eq:T2nd}), we choose $\cos(\omega t)$ at the initial position so as to obtain a real Floquet Hamiltonian, and yet different results will be obtained with different phases for the diabatic coupling, e.g. $\cos(\omega t + \zeta)$. Such results will be presented and discussed in a following paper.\cite{zeyu2019robust} For real life experiments, unless one can lock in a relationship between the initial phase of the diabatic coupling and the initial position of an initial wavepacket, one would need to average overall initial phases $\zeta$ (or several initial positions),  and presumably, the averaged results would be identical to the blue dashed line labeled as "F-FSSH pop". Thus, the question of whether or not we can observe the oscillation in transmission (as predicted in Figs.~\ref{fig:T1st_instantad_F-FSSH} -~\ref{fig:reflecs_2nd_instantad_F-FSSH}) remains open.

Lastly concerning the case $\omega\rightarrow 0$, F-FSSH cannot always guarantee reliable electronic amplitudes $\tilde{c}_{n\mu}$ for calculating the interference term in Eqs.~(\ref{eq:FFSSH_interf}) and~(\ref{eq:FFSSH_damp}), as shown in the $\omega = 0.00001$ case, especially when the intensity of the coupling is strong or the frequency is low enough to get more Floquet states involved. We conclude that the regime in which F-FSSH breaks down is when both the coupling strength $V$ and the kinetic energy of the nuclei $T$ are much greater than the energy $\hbar\omega$ of the light.
In such a case, for now, we believe the prescription of Ref. [\onlinecite{fiedlschuster2016floquet}] is good enough: run F-FSSH when $V \leq \hbar \omega$ and switch to standard FSSH when $V \gg \hbar \omega$.

In the end, F-FSSH clearly has promise for propagating dynamics with time-dependent couplings. The algorithm is stable, efficient and easily incorporated to deal with complicated problems involving light-matter interaction, especially at reasonably low intensity. In our calculations, the cost of F-FSSH is roughly six times the cost of IA-FSSH. Furthermore, we believe many improvements are possible in the future, with regards to coherence and decoherence. Assuming the algorithm can be extended easily and accurately to the case that as in Ref.~[{\onlinecite{fiedlschuster2016floquet}}], we expect F-FSSH to be the focus of much attention in the years to come.

\begin{acknowledgments}
This work was supported by the U.S. Air Force Office of Scientific
Research (USAFOSR) AFOSR Grants No. FA9550-18-1-0497 and FA9550-18-1-0420.
J.E.S. and A.N. thank Eli Pollak for very interesting conversations about Floquet theory.
\end{acknowledgments}

\nocite{*}
%
\providecommand{\noopsort}[1]{}\providecommand{\singleletter}[1]{#1}%
\end{document}